# Musical analysis of Stravinski's *The Rite of Spring* based on computational methods

German Ruiz Marcos[1]*

**Abstract**
Stravinski's *The Rite of Spring* is one of the most well-known pieces from the classical contemporary music repertoire. However, its analysis has aroused different opinions within its contstruction and compositional foundations. In this sense, I here proposed my own manual analysis and a computational approach which aims to find a similar analysis, giving the opportunity of discovering new possible points of view and supplying the current deficiencies of the Music Computing common analysis systems.

[1]*Student of the Sound and Music Computing Master Programme (Universitat Pompeu Fabra)*
*****Corresponding author**: german.ruiz02@estudiant.upf.edu

## Contents

## Introduction

According to Mikhailovich-Yarustovsky, "a new musical language has emerged in Stravinski's *The rite of spring*" [1]. Nowadays, this musical piece is considered as one of the most outstanding contributions to contemporary music, becoming a major influence of the 20th-century's leading composers. However, the original performance, that took place on 29 May 1913, had a mixed critical reception and the ballet was not performed again until the 1920s. The narrative behind The Rite of Spring is based on the sacrifice of a young girl who, after primitive rituals celebrating the advent of spring, dances herself to death [2].

The computational methods used at present in musical analysis are not correctly adapted and updated to the contemporary techniques that have been used since the late decades of the 19th century. That is to say, current computational ideas behind the musical analysis are based, somehow, on classical structure, steady use of key and tonality, major and minor scales or basic rhythmic patterns. However, the 20th century was the epoch of the cultural, social, economical and human revolution and music had to deal with its narrations, what translated into the construction of a flexible, personal and complex new musical system. So that, the computational approaches based on working with and within contemporary pieces need to be re-design in terms of each specific theoretical aspects, taking into account that some of the novelty that would appear on those pieces could just be explained in a theoretical framework based on perceptual and cognitive assumptions and not musical features.

Thereby, I decided to analyse myself by hand Stravinski's *The Rite of Spring* and compare my own analysis with a basic automatic one. As expected, all the output information I obtained did not make any sense. Thus, I decided to compute a basic analytical-system that would consider the concepts of tonality, key, scale, mode, tonal centre, rhythmic pattern and musical interval from a contemporary perspective. In this sense, I describe in Section 1 my own analysis of *The Rite of Spring*, as well as a summary of my computational approach based on a main python file and its util functions. After that, I provide in Sections 2 the results I have obtained during my research and some discussion based on the manual analysis, respectively. Finally, I end up with some conclusions.

# 1. Methodology

To start with, I decided to analyse the piano four-hands arrangement written by Stravinski himself. Likewise, I have just focussed on the *Introduction* movement of the Part I. *The Adoration of the Earth*; that is to say, the first three minutes of the piece. Then, I performed a musical analysis based on the contemporary theoretical frameworks as well as lots of listening annotations. Thereby, based on this previous findings, I computed a midi extractor where the data was segmented according to some melodic and harmonic guidelines. So that, after using specific functions related to notes, chords, figures and scales extraction, I acquired some interesting results that could help in creating a general idea of my analytical proposal of *The Rite of Spring*.

## 1.1 Manual Analysis

In my opinion, the *Introduction* movement can be split into 10 different sections, as described in Table 1, according to the analysis of parameters such as: tonal centre, intervals movements, variable tonality, steady tonality, important altered notes, strong beats and rhythmic grouping, among others. So that, I am now providing a summary of the characteristics of each section.

**Table 1.** Sections structure

| Section | Measures |
|---|---|
| A | 01 to 12 |
| B | 13 to 19 |
| C | 20 to 24 |
| D | 25 to 27 |
| E | 28 to 31 |
| F | 32 to 38 |
| G | 39 to 45 |
| H | 46 to 51 |
| I | 52 to 65 |
| A' | 66 to 68 |
| J | 69 to 75 |

In *Section A*, a predominant melody is introduced in aeolian A. It could be discussed whether the section is written in C major but, from my point of view, the A note plays a really important role. Furthermore, coming sections would stress E and B tonalities, what could be understood as a classical usage of dominant and dominant of the dominant functions. Then, the melody continues but with several changes by altering some notes. The alterations respond to the accompaniment, which moves in two-notes blocks, with a perfect fourth interval between notes, following a 'Circle of Fourths'; that is to say, after the first intervention in D major (DM) in the final cadence of Section A, the following chord would belong to a tonality a fourth away. However, just the pitch class of each tonality would match this rule, so that, every new event could be in fourth-distance scale in a sharp (#), flat (b), major (M) or minor (m) tonality. The harmonic scheme that corresponds to this section is: DM-AbM-EbM-BbM-FM-CbM-Fm-Bm-EM-Am-DM-Gm-DM-Am. Finally, the melodic idea returns combined with the previous harmonic idea, following a Em-BM-FM-EM structure. That is to say, this passage makes up a half cadence (VI-V of A), giving the sensation of finishing the introduction of the predominant melody. Likewise, an A major coda is added at the end. On the other hand, this section shapes an 8-measures classical phrase, responding to the typical question-answer structure, and stressing beats every periodic bar.

In *Section B*, a small excerpt of predominant melody is played over a secondary harmonic structure that follows again the "Circle of Fourths" structure by playing the sequence Fm-Cm-Fm-Bm-EM-Am. After that, a new harmonic succession takes the leading role based on a sequence of triplets that go through Am-D#m-G#m-Dm-AM-EM-Bm-F#M-CM-GM-DM-AM-EM-BM-F#m-C#m-GM-DM-AM-EM-AM-DM-GM-C#M-F#m-BM-EM-AM-DM-AM-EM-Bm.

In *Section C*, a steady tonality is reached and a new melodic theme is played over a simple BM harmony (A dominant's) filled with tremolos.

In *Section D*, the melody becomes complex and it is ornated with plenty of effects. Likewise, the harmony is based on alternative sequences of polychords of the tonic of AM and the dominant of D#m.

In *Section E*, the melody's complexity of the previous section moves to the bass accompaniment which is based on arpeggios and effects over Em while the melody plays some counterpoint in the octatonic scale (*diminished D*), creating a small polytonal passage.

In *Section F*, the musical content is reduced to triplets that follow a major third + perfect fourth structure. The harmonic structure is repeated twice based on patterns such as $I_6^{11} - III^{11} - IV_{64}^{13} - I_{64} - IV_{64}^{13} - III^{11} - V_{64}^{11}$, functions that correspond to the sequence of tonalities BM-BbM-Dm-BbM-GM-BbM.

In *Section G*, a question-answer structure is used, being the question melody in a C whole-tone scale accompanied by a sequence of triplets based on augmented 6th and minor 3rds intervals and, the respond, a set of contrapuntal excerpts in an octatonic scale.

In *Section H*, the melody begins to incorporate more instrumentation and complexity in a steady BM section accompanied by some tremolos, playing notes separated a three tones interval (diminished 5th or augmented 4th).

In *Section I*, the melody is split in different interventions over a Gm harmony filled with effects and density. After an introduction of the harmonic and melodic resources to be used, the bass plays an ostinato of $V\tilde{I}^{13} - I$ again in Gm but, using enharmonics.

In *Section A'*, the first predominant melody used in Section A is introduced again but in Abm.

In *Section J*, the movement concludes by creating a rhythmic pattern over 16th notes played with staccatos following a minor 3rd and perfect 4th pattern, concluding with an $A_7^{7_{11}}$; that is to say, an imperfect cadence that perfectly connects with the next movement.

## 1.2 Main Code

In order to carry out a musical analysis based on computational techniques, I decided to compute a simple method based on a really basic theoretical framework adapted to Stravinski's *The Rite of Spring*.

### 1.2.1 Midi extraction

The starting point would be to extract the midi information to work with. I then created a text file where basic information such as ONSET (BEATS), DURATION (BEATS), MIDI CHANNEL, MIDI PITCH, VELOCITY, ONSET (SEC) and DURATION (SEC) was stored from MIDI tool-box extraction. Likewise, some transformations need to be done to convert the MIDI tool-box [3] vertical writing style to a horizontal style, where each line corresponds to the same piece of midi-audio.

### 1.2.2 Segmentation and analysis

Once the midi information is written in an external text file, the predominant melody is extracted by annotating the midi notes of the channel that was observed to be the one containing it. In this specific case, we are talking about channel 5. Likewise, in order to compute a rhythmic and harmonic analysis, the simultaneous events are grouped in blocks by taking into account the existence of different channels within each event.

Finally, to analyse the piece, the previously commented event-blocks sections were used as segments in the chords extraction and the rhythmic pattern matching processes. However, in order to compute a melodic analysis, the predominant melody has to be split into small segments. My first approach was based on creating melodic lines taking into account the onset and duration information in seconds. So that, I created different melodic phrases by selecting a time threshold and separating the predominant melody where a pause/silence was detected. Nevertheless, the result obtained using this method did not really match a realistic segmentation, which is crucial for finding a tonal centre and a mode. Thereby, I computed the current method: the two first notes were annotated as a new melody and a set of possible tonal centres and modes was computed. Then, if those were correctly found, a new note was included into the new melody and the process was repeated until no tonal centre and mode could match the melody. In this way, the different phrases within the whole predominant melody were constructed according to the possibility of finding a potential tonal centre and mode.

## 1.3 Util functions

For the purpose of analysing *The Rite of Spring* by using my piece of code, a set of functions were needed.

### 1.3.1 Notes estimation

In order to correlate the midi notes with their notes names, the *fun_notes* takes a midi note number and checks whether it is contained in a set of pre-defined lists, as seen in Figure 1. So that, the function outputs a string with the label of the note as well as its midi value within the octave number 2, that is to say, between 24 and 35. The notes boundaries of *The Rite of the Spring* midi notes are [34:99] and, that is why the notes lists include seven different octaves.

$$\begin{aligned}
midiC &= (24, 36, 48, 60, 72, 84, 96) \\
midiCs &= (25, 37, 49, 61, 73, 85, 97) \\
midiD &= (26, 38, 50, 62, 74, 86, 98) \\
midiDs &= (27, 39, 51, 63, 75, 87, 99) \\
midiE &= (28, 40, 52, 64, 76, 88, 100) \\
midiF &= (29, 41, 53, 65, 77, 89, 101) \\
midiFs &= (30, 42, 54, 66, 78, 90, 102) \\
midiG &= (31, 43, 55, 67, 79, 91, 103) \\
midiGs &= (32, 44, 56, 68, 80, 92, 104) \\
midiA &= (33, 45, 57, 69, 81, 93, 105) \\
midiAs &= (34, 46, 58, 70, 82, 94, 106) \\
midiB &= (35, 47, 59, 71, 83, 95, 107)
\end{aligned}$$

**Figure 1.** Midi notes lists

### 1.3.2 Tonal Centre estimation

Thinking about different approaches with which estimate the tonal centre was not an easy task. Moreover, the accuracy of its estimation would translate into the possibility of finding the correct mode; on the contrary, missmatching the tonal centre would most probably mean an incorrect mode labelling. In this sense, I computed the *fun_tonal_centre* for labelling a certain melody.

**Most played note**   My first approach was based on considering the note that appears the most as the tonal centre. However, the existence of nonchord tones, flourishes and passing notes implied the possibility of mismatching the tonal centre when applying this technique, what surprisingly happened a lot. This method was used to compute the scale using *fun_scale*.

**Most frequent note**   My second approach was based on considering the most frequent note as the tonal centre. This was computed by annotating the duration of each note, reducing all the pitch classes into one single octave. However, although the results showed an improvement over the previous method, a similar problem was found: not always the most played note matches the tonal centre. It may happen that the sum of durations of a certain note is bigger than the tonal centre's total duration. This method was used to compute the scale using *fun_prov_scale*.

**Random candidates**   My final approach was influenced by the 'Key profiles' [4] estimation method. So that, a specific melody was transformed into a set of its corresponding pitches within just one octave. After that, a possible tonal centre was set as a candidate and it was checked if there existed possible modes that matched the melody sequence. The final selection of the tonal centre was done together with the mode labelling based on my own 'Mode profiles'. So that, this will be explained in the next function. This method was used to compute the scale using *fun_prof_scale*.

### 1.3.3 Scale estimation

Based on the tonal centre extraction, the mode was computed as the list that contains all the values within the melody. The modes were defined as it could be seen in Figure 2. When computing the tonal centre as the most played or frequent note, not always a possible scale was found for each set of notes. The method was really simple, based on just checking whether the melody notes' values were included in one of the mode lists.

$$
\begin{aligned}
ionian\_major &= [i, i+2, i+4, i+5, i+7, i+9, i+11] \\
dorian &= [i, i+2, i+3, i+5, i+7, i+9, i+10] \\
phrygian &= [i, i+1, i+3, i+5, i+7, i+8, i+10] \\
lydian &= [i, i+2, i+4, i+6, i+7, i+9, i+11] \\
mixolydian &= [i, i+2, i+4, i+5, i+7, i+9, i+10] \\
aeolian &= [i, i+2, i+3, i+5, i+7, i+8, i+10] \\
minor &= [i, i+2, i+3, i+5, i+7, i+8, i+10] \\
locrian &= [i, i+1, i+3, i+5, i+6, i+8, i+10] \\
whole\_tone &= [i, i+2, i+4, i+6, i+8, i+10] \\
octatonic1 &= [i, i+1, i+3, i+4, i+6, i+7, i+9, i+10] \\
octatonic2 &= [i, i+2, i+3, i+5, i+6, i+8, i+9, i+11]
\end{aligned}
$$

**Figure 2.** Mode lists for 'i' tonal centre

On the other hand, I finally computed a set of 'Mode profiles' based on the usage, meaning and correlation of the different functions of a specific mode, as seen in Table 2.

The selection of the weights of each mode was based on my personal assumptions of the typical uses, meanings and functions of each grade of the scales. In this sense, I took into account the following reasons for selecting those values:

* **Ionian**

**Table 2.** Modes profiles weights

| Mode | 5-value | 3-value | 1-value |
|---|---|---|---|
| Ionian | I IV V | III VI | II VII |
| Dorian | I V VII | III IV | II VI |
| Phrygian | I II V | III VI | IV VII |
| Lydian | I II V | IV VII | III VI |
| Mixolydian | I IV VII | V VI | II III |
| Aeolian | I III VI | V VII | II IV |
| Locrian | I II V | IV VII | III VI |
| Whole-tone | IV V VI | I | II III |
| Octatonic 1 | III V VI | I | II IV VII VIII |
| Octatonic 2 | II III V | VIII | I IV VI VII |

  i. *Value 5:* tonal functions
  ii. *Value 3:* modal functions
  iii. *Value 1:* the rest

* **Dorian**
  i. *Value 5:* most common chords
  ii. *Value 3:* dominant chords
  iii. *Value 1:* the rest

* **Phrygian**
  i. *Value 5:* most common chords
  ii. *Value 3:* distinction from major and minor
  iii. *Value 1:* the rest

* **Lydian**
  i. *Value 5:* all major triads
  ii. *Value 3:* diminished chords
  iii. *Value 1:* the rest

* **Mixolydian**
  i. *Value 5:* distinction from ionian
  ii. *Value 3:* most common chords
  iii. *Value 1:* the rest

* **Aeolian**
  i. *Value 5:* distinction from ionian
  ii. *Value 3:* most common chords
  iii. *Value 1:* the rest

* **Locrian**
  i. *Value 5:* distinction from minor
  ii. *Value 3:* tritone above the tonic
  iii. *Value 1:* the rest

* **Whole-tone**
    i. *Value 5:* altered differences
    ii. *Value 3:* tonic
    iii. *Value 1:* the rest
* **Octatonic 1**
    i. *Value 5:* altered differences
    ii. *Value 3:* tonic
    iii. *Value 1:* the rest
* **Octatonic 2**
    i. *Value 5:* altered differences
    ii. *Value 3:* most common chord
    iii. *Value 1:* the rest

So that, as previously commented, after segmenting the melody by setting the possible pair of candidates, *[tonal centre - mode]*, all the modes are weighted and compared with each other. Then, the mode highest weight for each mode profile of every melody would finally establish the tonal centre (predominant pitch class) and the mode.

In Figure 3 the modes profiles are shown in a graph.

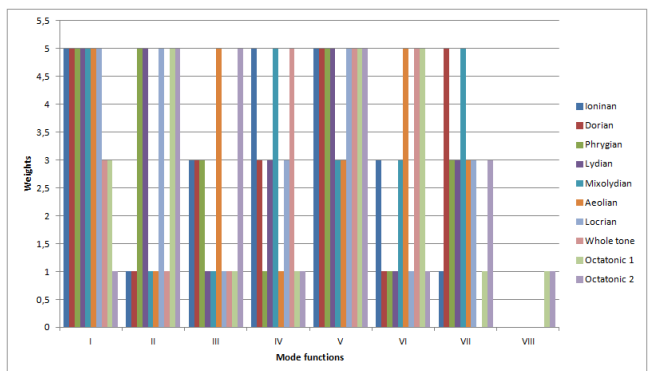

**Figure 3.** Modes profiles

### 1.3.4 Chords estimation

After segmenting the piece into blocks based on simultaneous events, a list of intervals and chords are used for labelling chords. So that, taking each segment (block), the events are analysed according to their channel; that is to say, the 'harmonic' analysis is just computed for notes that are played on the same channel at the same time. And, again, the notes within the segments are compared with the pre-defined lists in order to determine whether the set of events is included in those intervals/chords. The analysis is computed in a loop, thereby, the chords are ordered from the most specific to the least. This means that a chord could be included in several categories, *e.g.* a tonic and a major third could be part of a perfect chord, but also it could be a dominant 7th, a major 9th, a major 11th, etcetera. In this way, the final choice of a chord would be the most accurate one. For instance, a perfect major chord could be first classified as a perfect 13th, then major 11th, major 9th, dominant 7th and, finally, perfect major.

On the other hand, the chords inversions information is lost on purpose. The chords are analysed from the fundamental and they are not reconstructed after the analysis. This is because the goal of *fun_chords* is to give a general overview of the harmonic distribution and not a deep analysis. If a more concrete information is required, is is very probable that a whole new (and extremely complex) perspective has to be proposed for computing Stravinski's analysis.

### 1.3.5 Rhythm estimation

Finally, focussing on rhythm was quite a challenge. There are many aspects in Stravisnki's composing style related to rhythm and, in special, in *The Rite of Spring*. Furthermore, the most common analysis that could be done related to rhythm, also the one I proposed myself in my manual analysis, are based on structure, segmentation, tempo, meter and beats strength. However, most of those ideas are related to music perception and cognition and are just accessible from symbolic data and musical features extracted from audio. Thereby, I decided to work on analysing polyrhythm by channels in the same way I have done with harmony (chords). Taking advantage of the simultaneous events segmentation, I used *fun_polyrhythm* to find correlations between the durations of the events within every simultaneous event. Therefore, I transformed the onset and duration beat information into figures annotations, establishing polyrhythmic patterns. However, the analysis window is just taken as the location where the simultaneity takes places, so, for those notes that were played before that instant, or that continue after it finishes, the polyrhythm would be found but with a mislabeled value of those notes. Nevertheless, the obtained information could give a general idea of the distribution of rhythmic patterns between channels.

# 2. Results and Discussion

By computing the main code, where the util functions extract information from the midi data, a melodic, harmonic and rhythmic analysis is obtained as output.

## 2.1 Melodic analysis

When the tonal centre is computed as the most frequent note of the analysed melody, the output corresponds to Figure 4. Nevertheless, this first approach seems to be a bit precarious. Most of the tonal centres are mislabelled what translates into wrong modes classification. This first look to the melodic perspective tells us that the technique needs to be improved.

When the tonal centre is set as provisional within the chromatic scale in order to detect which are the possible modes and one is randomly selected, the output corresponds to Figure 5. In this case, at least all the segments have been assigned with a mode. However, the tonal centre selection is no accurate enough, what gives some weird outputs when talking about the scales.

When the tonal centre and mode are set as all the possible candidates of a certain melody and the final decision is based on the previously introduced *Mode Profiles*, the output corresponds to Figure 6 for the 'Predominant Melody' (*channel 5*), and Figure 7 for the 'Secondary melodies' (*channel 4*). So that, 6 is the final approach of the melodic analysis. In the first segment, it could be noted that a possible tonal centre-mode is C major, what makes sense if we take into account that the notes are the same as in the A aeolian scale, the one I have suggested as the correct one. However, this possibility is discarded. This could through some light into the design of the Modes Profiles, showing a problem in the lydian weight's decision.

The next segment is labelled as octatonic 1 but it corresponds to a polytonal sequence. However, the notes of the predominant melody are the same as the ones within the C-octatonic 1 scale, what makes sense. Moreover, if we take a look into 7, the 'Secondary Melody', it could be seen that every time a new event is included in the melody segment, a new tonal centre and mode is identified. That is to say, tonality is not stable in a passage filled with accidental alterations. This evidences the existance of the polytonal sequence found in the manual analysis.

The third segments behaves exactly the same as the first segment. So that, same problems.

In the fourth segment, a really big piece of music has been analysed. Melodically speaking, the piece has gone through a B major section, a polychordal sequence and an octatonic scale melody. So that, there has been a segmentation problem and, therefore, the selected tonal centre and mode are not reliable.

To continue, in segment number five, the classification seems to be wrong but, again, the actual C-whole predominant melody shares alterations with the predicted D major sequence and, additionally, the predominant melody does not play the alterations that could differentiate both scales.

Finally, in the sixth section, the selected scales is two semitones far away from the correct analysis. However, the computational analysis recognises the enharmonization change made in the middle of that section, what supports my personal annotations.

To conclude, the analysis of the predominant melody finishes here. However, the music continues in the piece. This is because it is the secondary melody the one that takes the leading role from here to the end, matching again a possible C as tonal centre, but evidencing one more time the probable mistake in the lydian 'Modes Profiles' weights.

## 2.2 Harmonic analysis

Likewise, the harmonic analysis is computed over the simultaneous events locations, where the output corresponds to Figure 8. The harmonic analysis results show a clear failure of the proposal. It has to be taken into account that the proposed approach just focusses on intervals within the same channel. However, the majority of the harmonic constructions of the piece are based on the combination of the intervals of different channels, what gives the foundations of the chords constructions. So that, the analysis fails from Sections A to H. After that, it finds a perfect fourths sequence in the Section I introduction. Likewise, it matches the bass ostinato that starts in the middle of section I by labelling the dominant 7th and the perfect chords.

## 2.3 Rhythmic analysis

Finally, the rhythmic patterns detected within the simultaneous events conrrespond to Figure 9. From the rhythmic analysis, it could be said that the extracted information matches, in general, the behaviour of the detected simultaneous segments. However, the output information given is again just based on equal channel events emergences. Likewise, the windowing issue seems to not have been well settled. Therefore, the final analysis barely gives the chance of creating rhythmic scheme or mental image of the piece. On the other hand, it has to be said that, even having detected correct polyrhythmic sequences, analysing rhythm, and even more when talking about Stravinski and *The Rite of Spring*, is a hugely complex task that would need from wide analysis of beat tracking, rhythmic patterns and strong beats location. That is to say, maybe the needed approach for analysing rhythm was not a realistic option in this piece of research.

# 3. Conclusions

To sum up, this paper has proposed a computational approach for analysing Stravinski's *The Rite of Spring*. By taking a look at the results, it can be argued that the melodic analysis here introduced is not far away from the optimal method to be used, taking into account several ideas, previously described, that would help in improving the code. However, some light has been thrown into the 'Modes Profiles' proposal, evidencing its strengths and weaknesses.

On the other hand, the harmonic and rhythmic systems here used are still far away from an admissible analysis technique. Moreover, the main problems related to these methods have been detected and highlighted but, the obviousness of how expensive could be to incorporate the needed improvements has more weight on the scale when thinking about the utility of the implementation; that is to say, is it possible to continue working on this Stravinski's specific analysis but with the certainty that the effort and the work employed will be useful in the future for different projects? Or put another way: Should it be the computational analysis of contemporary music specific to each composer or even worse, specific to each piece? ———————————————————————
———————

## References

[1] I. Stravinsky. The rite of spring. courier corporation. 2000.

[2] J. Gittelsohn. The rite of spring. scholastic update, 124(13), 14-16. 1992.

[3] T. Eerola and P. Toiviainen. Midi toolbox: Matlab tools for music research. 2004.

[4] Carol L. Krumhansl and E. J. Kessler. Tracing the dynamic changes in perceived tonal organisation in a spatial representation of musical keys. psychological review, 89:334–368. 1982.

**Onset time/Scale/Mode:**

| | | | |
|---|---|---|---|
| 0.0000 | 17.1636 | B/Cb | notknownmode |
| 17.2309 | 24.5303 | C/# | octatonic1 |
| 26.2450 | 46.5295 | B/Cb | notknownmode |
| 57.2870 | 100.0143 | E/Fb | octatonic1 |
| 100.2567 | 107.2113 | G | notknownmode |
| 129.6431 | 150.9113 | D | notknownmode |

**Figure 4.** Melodic analysis 1st trial (tonal centre = most frequent note)

**Onset time/Scale/Mode:**

| | | | |
|---|---|---|---|
| 0.0 | 17.1636 | C/B# | lydian |
| 17.2309 | 24.5303 | C/B# | octatonic1 |
| 26.245 | 46.5295 | C/B# | lydian |
| 57.287 | 100.0143 | C#/Db | octatonic1 |
| 100.2567 | 107.2113 | D | mixolydian |
| 129.6431 | 150.9113 | C#/Db | octatonic1 |

**Figure 5.** Melodic analysis 2nd trial (tonal centre = random possible)

**Onset time/Scale/Mode/Pair candidates/Respective Weights:**

| | | | | | |
|---|---|---|---|---|---|
| 0.0 | 17.1636 | C/B# | lydian | [[C/B#$^{0,0}$ionian_major$^0$], [C/B#$^{0,0}$lydian$^0$]] | [113, 133] |
| 17.2309 | 24.5303 | C/B# | octatonic1 | [[C/B#$^{0,0}$octatonic1$^0$]] | [18] |
| 26.245 | 46.5295 | C/B# | lydian | [[C/B#$^{0,0}$ionian_major$^0$], [C/B#$^{0,0}$lydian$^0$]] | [104, 116] |
| 57.287 | 100.0143 | C#/Db | octatonic1 | [[C#/Db$^{0,0}$octatonic1$^0$], [E/Fb$^{0,0}$octatonic1$^0$]] | [16, 14] |
| 100.2567 | 107.2113 | D | ionian_major | [[C/B#$^{0,0}$lydian$^0$], [C#/Db$^{0,0}$locrian$^0$], [D$^{0,0}$ionian_major$^0$], [D$^{0,0}$mixolydian$^0$]] | [30, 30, 38, 32] |
| 129.6431 | 150.9113 | C#/Db | octatonic1 | [[C#/Db$^{0,0}$octatonic1$^0$]] | [115] |

**Figure 6.** Predominant melody analysis - CHANNEL 5 - (Mode Profiles)

**Onset time/Scale/Mode/Pair candidates/Respective Weights:**

15.3078   20.3078 C#/Db   octatonic1   [[$C#/Db^{0,0}octatonic1^0$], [$E/Fb^{0,0}octatonic1^0$]]   [15, 9]

23.0001   29.0912 D   octatonic1   [[$C#/Db^{0,0}phrygian^0$], [$C#/Db^{0,0}aeolian^0$], [$C#/Db^{0,0}minor^0$], [$D^{0,0}lydian^0$], [$D^{0,0}octatonic1^0$], [$D#/Eb^{0,0}locrian^0$], [$E/Fb^{0,0}ionian_major^0$], [$E/Fb^{0,0}mixolydian^0$], [$F/E#^{0,0}octatonic1^0$], [$F#/Gb^{0,0}dorian^0$], [$F#/Gb^{0,0}aeolian^0$], [$F#/Gb^{0,0}minor^0$], [$G#/Ab^{0,0}phrygian^0$], [$G#/Ab^{0,0}locrian^0$], [$G#/Ab^{0,0}octatonic1^0$]]   [11, 13, 7, 13, 15, 13, 13, 11, 7, 7, 11, 7, 15, 15, 5]

29.4758   63.6355 C#/Db   octatonic1   [[$C#/Db^{0,0}octatonic1^0$]]   [306]

65.9234   69.0295 C/B#   aeolian   [[$C/B#^{0,0}aeolian^0$], [$C/B#^{0,0}minor^0$], [$D^{0,0}locrian^0$]]   [212, 180, 162]

74.1052   77.4386 D   ionian_major   [[$C/B#^{0,0}lydian^0$], [$C#/Db^{0,0}locrian^0$], [$D^{0,0}ionian_major^0$], [$D^{0,0}mixolydian^0$]]   [36, 36, 40, 34]

80.2795   123.3477   D   aeolian   [[$C/B#^{0,0}mixolydian^0$], [$C#/Db^{0,0}octatonic1^0$], [$D^{0,0}aeolian^0$], [$D^{0,0}minor^0$]]   [61, 47, 115, 53]

123.5749   136.7113   C/B#   dorian   [[$C/B#^{0,0}dorian^0$], [$C/B#^{0,0}aeolian^0$], [$C/B#^{0,0}minor^0$]]   [134, 118, 134]

136.7795   139.7719   C#/Db   octatonic1   [[$C#/Db^{0,0}octatonic1^0$]]   [4]

139.8931   142.7492   C/B#   minor   [[$C/B#^{0,0}aeolian^0$], [$C/B#^{0,0}minor^0$]]   [45, 49]

147.6658   168.511   C/B#   lydian   [[$C/B#^{0,0}ionian_major^0$], [$C/B#^{0,0}lydian^0$]]   [327, 333]

**Figure 7.** Secondary melody analysis - CHANNEL 4 - (Mode Profiles)

**Fundamental/Chord/Time onset/Midi channel:**

*G#/Ab*   *P4*     58.3477     13.0000
*D#/Eb*   13*th*    58.5749     13.0000
*F/E#*    13*th*    73.9158     4.0000
*D#/Eb*   13*th*    95.6204     2.0000
*B/Cb*    *P4*     130.0143    13.0000
*B/Cb*    *P4*     130.6204    13.0000
*B/Cb*    *P4*     131.2264    13.0000
*B/Cb*    *P4*     131.8325    13.0000
*B/Cb*    *P4*     132.4386    13.0000
*B/Cb*    *P4*     133.0446    13.0000
*B/Cb*    *P4*     133.6507    13.0000
*B/Cb*    *P4*     134.2568    13.0000
*B/Cb*    *P4*     134.8628    13.0000
*B/Cb*    *P4*     135.4689    13.0000
*G#/Ab*   13*th*   136.5295    4.0000
*B/Cb*    *P4*     136.681     13.0000
*F/E#*    13*th*   136.7113    4.0000
*D#/Eb*   13*th*   136.8401    4.0000
*D*       13*th*   136.9007    2.0000
*B/Cb*    *P4*     137.2871    13.0000
*B/Cb*    *P4*     137.8931    13.0000
*B/Cb*    *P4*     138.4992    13.0000
*B/Cb*    *P4*     139.1052    13.0000
*G*       13*th*   139.7719    4.0000
*D*       13*th*   140.0825    2.0000
*G#/Ab*   *dom*7   144.0295    0
*G#/Ab*   *dom*7   144.484     0
*G*       13*th*   144.7719    2.0000
*G#/Ab*   *dom*7   145.3931    0
*E/Fb*    *A*4     145.9234    0
*D*       13*th*   146.9007    2.0000
*G*       13*th*   148.8628    2.0000
*D*       13*th*   149.1734    2.0000
*E/Fb*    *A*4     149.5598    0
*F/E#*    *A*6     149.6204    1.0000

**Figure 8.** Harmonic analysis (intervals and chords)

**Event(sec)/Figures-polyrhythm by channel:**

[59.7113, [$\phi$*tri plethal f*$^{0}$,$^{0}$16*th f lourish*$^{q}$]]]
[60.6204, [$\phi$[16*th f lourish*$^{0}$,$^{0}$*triplet*16*th* $^{q}$]]]
[61.0749, [$\phi$[16*th f lourish*$^{0}$,$^{0}$*triplet*16*th* $^{q}$]]]
[64.5598, [$\phi$*tri plet*8*th*$^{0}$,$^{0}$*triplet*16*th* $^{q}$]]]
[66.0749, [$\phi$*quarternote*$^{0}$,$^{0}$16*th f lourish*$^{q}$]]]
[66.984, [$\phi$*triplet*16*th* $^{0}$,$^{0}$16*th f lourish*$^{0}$,$^{0}$*triplet*16*th* $^{q}$]]]
[67.4386, [$\phi$*triplet*16*th* $^{0}$,$^{0}$16*th f lourish*$^{0}$,$^{0}$*triplet*16*th* $^{q}$]]]
[67.5901, [$\phi$*triplet*16*th* $^{0}$,$^{0}$*tri plet*8*th*$^{q}$]]]
[67.8931, [$\phi$*triplet*16*th* $^{0}$,$^{0}$16*th f lourish*$^{0}$,$^{0}$*triplet*16*th* $^{q}$]]]
[68.3477, [$\phi$*triplet*16*th* $^{0}$,$^{0}$16*th f lourish*$^{0}$,$^{0}$*triplet*16*th* $^{q}$]]]
[68.4992, [$\phi$*triplet*16*th* $^{0}$,$^{0}$*tri plet*8*th*$^{q}$]]]
[68.8022, [$\phi$*triplet*16*th* $^{0}$,$^{0}$16*th f lourish*$^{0}$,$^{0}$16*th f lourish*$^{q}$]]]
[69.7113, [$\phi$*tri plet*8*th*$^{0}$,$^{0}$16*th f lourish*$^{q}$]]]
[70.1658, [$\phi$*triplet*16*th* $^{0}$,$^{0}$16*th f lourish*$^{q}$]]]
[71.0749, [$\phi$*triplet*16*th* $^{0}$,$^{0}$16*th f lourish*$^{q}$]]]
[94.2567, [$\phi$8*thnote*$^{0}$,$^{0}$*tri plet*8*th*$^{q}$]]]
[96.0749, [$\phi$*tri plet*8*th*$^{0}$,$^{0}$16*th f lourish*$^{q}$]]]
[96.984, [$\phi$*tri plethal f*$^{0}$,$^{0}$16*th f lourish*$^{q}$]]]
[98.8022, [$\phi$*tri plet*8*th*$^{0}$,$^{0}$16*th f lourish*$^{q}$]]]
[101.5295, [$\phi$*tri plet*8*th*$^{0}$,$^{0}$16*th f lourish*$^{q}$]]]
[102.4386, [$\phi$*tri plet*8*th*$^{0}$,$^{0}$16*th f lourish*$^{q}$]]]
[106.984, [$\phi$*tri plet*8*th*$^{0}$,$^{0}$16*thnote*$^{q}$]]]
[112.4386, [$\phi$*quarternote*$^{0}$,$^{0}$8*thnote*$^{q}$]]]
[112.4386, [$\phi$*tri plet*8*th*$^{0}$,$^{0}$16*thnote*$^{q}$]]]
[136.0749, [$\phi$16*thnote*$^{0}$,$^{0}$*triplet*16*th* $^{q}$]]]
[139.7113, [$\phi$[16*th f lourish*$^{0}$,$^{0}$16*th f lourish*$^{0}$,$^{0}$*triplet*16*th* $^{q}$]]]
[140.1658, [$\phi$8*thnote*$^{0}$,$^{0}$*triplet*16*th* $^{q}$]]]
[144.7113, [$\phi$[16*th f lourish*$^{0}$,$^{0}$8*thnote*$^{q}$]]]
[145.1658, [$\phi$8*thnote*$^{0}$,$^{0}$16*thnote*$^{q}$]]]
[146.0749, [$\phi$[16*th f lourish*$^{0}$,$^{0}$8*thnote*$^{q}$]]]

**Figure 9.** Polyrhythmical analysis